\documentclass[11pt,a4paper,onecolumn,reqno]{amsart}
\usepackage[a4paper, margin=2.3cm, bmargin=3cm]{geometry}

\usepackage{graphicx,stfloats} \usepackage{color}
\usepackage{amsmath}
\usepackage{amsaddr}
\usepackage{bm, soul}
\usepackage{mathtools}
\usepackage[colorinlistoftodos,prependcaption,textsize=tiny]{todonotes}
\usepackage{braket}
\usepackage[square,comma,sort&compress, numbers]{natbib}
\usepackage{enumerate}

\usepackage{tikz}
\usepackage{pgfplots}
\pgfplotsset{compat=1.9}
\usetikzlibrary{calc,arrows,fadings,decorations.pathreplacing,decorations.markings,patterns,shapes.geometric}
\tikzset{>=stealth,inner sep=0pt, outer sep=2pt,}
\tikzset{vecteur/.style={->,thick,color=black,smooth}}

\usepackage{placeins}

\renewcommand{\st}[1]{}

\usepackage[version=3]{mhchem} \usepackage{siunitx}
 \usepackage{multirow}

\usepackage{etoolbox}
\patchcmd{\pprintMaketitle}
  {\hrule\vskip12pt}
 {\hrule\vskip12pt\ifvoid\extrainfobox\else\unvbox\extrainfobox\par\vskip12pt\fi}
 {}{}

\newsavebox\extrainfobox

\usepackage{caption}
\title{Combined effects of heat loss and curvature on turbulent flame-wall interaction in a premixed dimethyl ether/air flame}

\author[stfs]{Driss Kaddar$^{1,*}$, Matthias Steinhausen$^{1}$, Thorsten Zirwes$^{2, 3}$, Henning Bockhorn$^{2}$, Christian Hasse$^{1}$, Federica Ferraro$^{1}$}
\email{kaddar@stfs.tu-darmstadt.de} 
\address[]{$^1$Technical University of Darmstadt, Department of Mechanical Engineering, Simulation of reactive Thermo-Fluid Systems, Otto-Berndt-Stra{\ss}e 2, 64287 Darmstadt, Germany\\
$^2$Engler-Bunte-Institute, Karlsruhe Institute of Technology, Engler-Bunte-Ring 7, 76131 Karlsruhe, Germany\\
$^3$Steinbuch Centre for Computing, Karlsruhe Institute of Technology, Hermann-von-Helmholtz-Platz 1, 76344 Eggenstein-Leopoldshafen, Germany
}

\begin{document}
\pagestyle{plain}

\begin{abstract} 
This study investigates the effects of curvature on the local heat release rate and mixture fraction during turbulent flame-wall interaction of a lean dimethyl ether/air flame using a fully resolved simulation with a reduced skeletal chemical reaction mechanism and mixture-averaged transport. 
The region in which turbulent flame-wall interaction affects the flame is found to be restricted to a wall distance less than twice the laminar flame thickness. 
In regions without heat losses, heat release rate and curvature, as well as mixture fraction and curvature, are negatively correlated, which is in accordance with experimental findings. 
Flame-wall interaction alters the correlation between heat release rate and curvature.
An inversion in the sign of the correlation from negative to positive is observed as the flame starts to experience heat losses to the wall. The correlation between mixture fraction and curvature, however, is unaffected by flame-wall interactions and remains negative. Similarly to experimental findings, the investigated turbulent side-wall quenching flame shows both head-on quenching and side-wall quenching-like behavior. The different quenching events are associated with different curvature values in the near-wall region. Furthermore, for medium heat loss, the correlations between heat release rate and curvature are sensitive to the quenching scenario.
\end{abstract}


\maketitle

\section{Introduction} \addvspace{10pt}
Recently, synthetic (oxygenated) fuels such as dimethyl ether (DME), which have a reduced soot and $\mathrm{NO_x}$ formation~\cite{Lumpp2011}, have been investigated as promising alternative diesel surrogates. 
To better understand the behaviour of these fuels, DME can serve as a starting point for systematic investigations in complex flow configurations. 
For DME combustion, flame speeds have been reported to depend strongly on stretch in terms of Markstein lengths~\cite{Chen2007}. Stretch, consisting of strain and curvature, has a significant effect on local flame properties, such as the flame speed and the local heat release rate (HRR), through the coupled effects of unequal heat and mass diffusion~\cite{Poinsot2005}.
In turbulent flows, the flame is subjected to unsteady stretch caused by turbulent flame wrinkling. The effect of stretch on the flame has been investigated in several numerical studies of turbulent premixed flames~\cite{Chen2000, Chakraborty2005, Aspden2016, Cecere2016}. Chakraborty et al.~\cite{Chakraborty2005} investigated the influence of differential diffusion on the local displacement speed in relation to flame curvature in statistically planar turbulent flames. Using a single-step chemical reaction and prescribing different Lewis numbers, they showed that correlations between the components of the displacement speed and curvature vary between Lewis numbers above and below unity.
In three-dimensional hydrogen-enriched methane/air slot flames, Cecere et al.~\cite{Cecere2016} show that local equivalence ratios and displacement speeds depend on curvature.

In practical combustion systems, the flame behaviour is additionally affected by flame-wall interaction (FWI). In the vicinity of the combustor walls, the flame is prone to heat losses that result in incomplete combustion, with a consequent reduction in the overall efficiency and an increase in pollutant formation~\cite{Poinsot2005}. During turbulent FWI, the flame is not only affected by heat losses, but also by turbulence-induced effects, such as flame stretch. 
This introduces a wide range of length and time scales, increasing the complexity of the phenomena in the near-wall regions. 
To further improve efficiency and reduce pollutant formation, a better understanding of the combined effects of turbulence and heat losses in near-wall flames is needed.
Therefore, turbulent side-wall quenching (SWQ) of premixed flames has recently been the subject of extensive experimental~\cite{Jainski2017, Kosaka2020, Zentgraf2021} and numerical~\cite{Gruber2010, Lai2019, Steinhausen2021, Ahmed2021, Wang2021, Jiang2021, Zirwes2021} studies investigating different aspects of the turbulent near-wall reacting flow. 
Gruber et al.~\cite{Gruber2010} studied a turbulent hydrogen/air flame with a direct numerical simulation (DNS), finding that near-wall coherent turbulent structures contribute significantly to the wall heat fluxes by pushing hot exhaust gas towards the wall. Further, they pointed out the importance of flame thickening during FWI.
In addition to hydrogen flames, multiple numerical studies of methane/air flames have been conducted~\cite{Jiang2021, Ahmed2021, Steinhausen2021}.
Jiang et al.~\cite{Jiang2021} studied the effects of FWI on the formation of \ce{CO} in a turbulent methane/air flame diluted with hot combustion products.
They found that the combustion process can be described by an ensemble of one-dimensional flames with differing strain rates and initial temperatures up to a wall distance of 0.25~mm. Closer to the wall, \ce{CO} formation is dominated by diffusion and convection, which underlines the challenges of FWI in terms of modelling.
Ahmed et al.~\cite{Ahmed2021} investigated the displacement speed and its components, as well as the strain rate and curvature statistics and their effect on the surface density function in a V-shaped flame ignited in a turbulent channel flow employing both adiabatic and isothermal wall boundary conditions. In their study, a single-step chemical reaction was employed and the Lewis number was assumed to be unity.
They found that close to the wall, the strain rate and flame curvature statistics exhibit differences depending on the wall boundary condition. Additionally, they proposed that the effects of varying the Lewis number should be investigated in future studies.
In addition to the numerical investigations, Zentgraf et al. conducted a comprehensive analysis of the thermochemical states in laminar~\cite{Zentgraf2022} and turbulent~\cite{Zentgraf2021} DME/air flames undergoing SWQ. The thermochemical state was characterized using simultaneous quantitative measurements of \ce{CO}, \ce{CO2} and temperature. In the laminar flame~\cite{Zentgraf2022}, the importance of differential diffusion effects during the FWI of DME/air flames was outlined. In the same burner configuration, Kosaka et al.~\cite{Kosaka2020} performed multi-parameter laser diagnostics of lean and stochiometric turbulent premixed methane/air and DME/air flames. In the study, the local HRR was estimated using a correlation based on normalized \ce{OH} and \ce{CH2O} planar laser-induced fluorescence measurements. The experimental data showed a negative correlation between the local HRR and the flame curvature. This again indicates a strong dependency on differential diffusion effects in DME/air flames during FWI.

To the authors' knowledge, a corresponding numerical study on the combined effects of flame stretch and heat losses on the local HRR during the turbulent FWI of DME/air flames has not been reported in the literature yet. The present work aims to close this gap by  investigating a lean DME/air flame ignited in a fully developed turbulent channel flow undergoing SWQ. A three-dimensional fully resolved numerical simulation is performed with reduced skeletal chemical kinetics. A mixture-averaged transport model is employed that allows the assessment of differential diffusion effects on the flame properties. 
The flame's equivalence ratio is similar to the experimental study by Kosaka et al.~\cite{Kosaka2020}. The main objectives of this work are: 
(i) to study how curvature affects the local HRR and mixture inhomogeneities caused by differential diffusion; 
(ii) to outline how heat losses in the near-wall region influence the local HRR and mixture inhomogeneities and their dependence  on curvature;
(iii) to investigate how local quenching modes affect the curvature statistics and the local HRR.
This work thus contributes to the understanding of FWI while adding additional physical constraints to the development of predictive combustion models for turbulent FWI in Large Eddy Simulations.

\section{Numerical setup} 
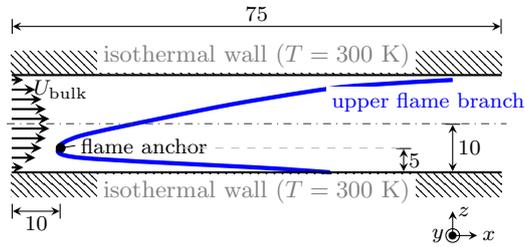
\begin{figure}[!htb]
    \centering
    \begin{tikzpicture} [scale=6.77/10.5]
     	\definecolor{azure}{rgb}{0.0, 0.5, 1.0}
    	\foreach \y in {-0.9,-0.7,-0.5,-0.3,-0.1,0.9,0.7,0.5,0.3,0.1}
    	\draw[vecteur] (0,\y)--++({2*(1-(\y*\y)^(1/2))^(1/7)-1.2},0);
        \draw (1,0.7)node{\scriptsize $U_\text{bulk}$};
    
        \draw[|<->|] (0,2.0)--++(10,0) node[midway, above]{\scriptsize 75};
        \draw[|<->|] (0,-1.8)--++(1,0) node[midway, below]{\scriptsize 10};
        
        \draw[ultra thick, blue] plot [smooth] coordinates { (9,0.9) (7, 0.78) (6, 0.65) (4.5, 0.4) (3, 0.1) (1.4, -0.25) (0.95, -0.5) (1.4, -0.7) (5, -0.9) (5.8, -0.95) (6.5, -1) };
        \node[blue, inner sep=2pt, fill=white, anchor=center] at (8.5, 0.45) {\scriptsize upper flame branch};
    
    	\filldraw (1,-0.5) circle (0.09);
        \draw[-] (1, -0.5) -- (1.3,-0.45) node[right]{\scriptsize flame anchor};
        \draw[|<->|] (9,-1)--++(0,1) node[midway, right]{\scriptsize 10};
        \draw[|<->|] (8,-1)--++(0,0.5) node[midway, right]{\scriptsize 5};
        \draw[lightgray, dashed] (3.5, -0.5)--++(4.5,0) node[right]{};

        \draw[->] (9.0,-2.3)--++(0,0.5) node[above, right]{\scriptsize $z$};
        \draw[->] (9.0,-2.3)--++(0.5,0) node[above, right]{\scriptsize $x$};
        \node[anchor=center,draw,circle] at (9.0, -2.3) {\tiny \textbullet};
        \node[anchor=center] at (8.7, -2.3) {\scriptsize $y$};

        \draw[thick](0,-1)--++(10,0);
    	\fill[pattern=north west lines] (0,-1.5) rectangle (10,-1);
    	\draw[thick](0,1)--++(10,0);
    	\fill[pattern=north west lines] (0,1) rectangle (10,1.5);
    	
        \draw[darkgray, dash dot] (-0.1,0)--++(10.2,0) node[right]{};

        \node[gray, inner sep=2pt, fill=white, anchor=center] at (5, -1.4) {\footnotesize isothermal wall ($T=300$ K)};
        \node[gray, inner sep=2pt, fill=white, anchor=center] at (5, 1.4) {\footnotesize isothermal wall ($T=300$ K)};
    
    \end{tikzpicture}
    \vspace{1 pt}
   \caption{Graphical illustration of the numerical setup in a lateral slice. The size of the domain in the lateral direction is 30~mm (three times the channel half-width). All measurements are given in mm.}
  \label{fig:DNS-setup}
\end{figure}

\addvspace{10pt}
The generic turbulent SWQ case analyzed in this work is inspired by the studies carried out by Gruber et al.~\cite{Gruber2010} and Steinhausen et al.~\cite{Steinhausen2021}. A V-shaped premixed DME/air flame is stabilized in a fully developed turbulent channel flow, where it undergoes side-wall quenching at the isothermal channel walls ($T_\text{wall}=300~\text{K}$). 
The inert channel flow has a frictional Reynolds number of $Re_\tau=\left( U_\tau H \right) / \nu = 180$, with $H$ being the channel half-width (10~mm), $U_\tau$ the wall shear velocity and $\nu$ the kinematic viscosity of the DME/air mixture at the inflow temperature (300~K). 
The level of the inlet turbulence
is characterized by the normalized velocity fluctuation $u^\prime/s_L = 0.76 $ with $u^\prime$ being the root mean square of the turbulent velocity fluctuations and the laminar flame speed $s_L$. The normalized integral length scale is $L_t/\delta_L = 25$ with the integral length scale $L_t$ and the thermal laminar flame thickness $\delta_L$. The thermal flame thickness is calculated by $\delta_L=(T_b - T_u)/max(dT/dx)$. $T_b$ and $T_u$ refer to the burnt and unburnt mixture temperatures, respectively. Following the definitions by Borghi and Peters~\cite{Borghi1985, Peters1988}, the flame is classified as inside the wrinkled flamelet regime.
Similar to Gruber et al.~\cite{Gruber2010}, the flame anchor is modeled as a cylindrical region ($r=0.9~\text{mm}$) of burnt gas temperature and composition. The inflowing DME/air mixture has an equivalence ratio of 0.83
and a Lewis number of $Le=1.6$ based on the deficient species.
A graphical illustration of the numerical configuration is shown in Fig.~\ref{fig:DNS-setup}.

The numerical simulation performed consists of two parts: an inert case and a reactive case:
\begin{itemize}
    \item \textbf{Inert case:} The inert simulation is used to produce suitable turbulent inflow conditions for the reactive simulation. The channel dimensions are $\left( x\times y\times z \right) = \left( 140 \times 30 \times 20 \right)~\text{mm}$, with $x$ being the stream-wise direction, $y$ the statistical independent lateral direction and $z$ the wall-normal direction. The channel walls are modelled with a no-slip boundary condition, while periodic boundary conditions are applied in both the lateral and stream-wise direction. The grid is refined towards the wall. In the core-flow, the non-dimensional grid spacing in the stream-wise, lateral and wall-normal direction is $\Delta x^+,~ \Delta y^+,~ \Delta z^+ = 2.5$. The superscript $+$ refers to a non-dimensionalization with the viscous length scale $\delta = u_{\tau}/\nu$, where $u_{\tau}$ is the wall friction velocity. The grid is refined towards the wall with a grid spacing of $\Delta z^+ = 0.4$ at the cell closest to the wall, resulting in a total of 61 million hexahedral cells.
    The initial turbulence field was generated by superimposing random fluctuations to the laminar mean flow profile with spatial and temporal correlation. The generated fluctuations are divergence-free and the implementation follows the work by Davidson~\cite{Davidson2007}. Starting from the initial turbulence field, the flow was simulated for ten flow through times to obtain a fully developed turbulent channel flow before sampling the velocity data for the reactive case. Additionally, the results have been benchmarked against a reference~\cite{Kim1987}.
    \item \textbf{Reactive case:} The channel width and height in the reactive simulation match these of the inert counterpart. The channel length is reduced to $75~\text{mm}$. The non-dimensional grid spacing in both stream-wise and lateral directions is $\Delta x^+,~ \Delta y^+ = 1.1$. This grid is also refined towards the bottom wall with a non-dimensional grid resolution of $\Delta z^+ = 0.27$ at the wall to ensure that the flame-wall interaction zone has a sufficient resolution~\cite{Steinhausen2021}. In total, the grid consists of 210 million hexahedral cells. 
    The flame front is resolved on a minimum of eight grid points.
    The inflow velocity fields in the inert case are stored with a time-step size of 5~\textmu s and are set as inflow conditions for the reactive case (interpolated spatially and temporally at every time step of the reactive simulation). The outlet is modelled using a zero-gradient boundary condition for the species, enthalpy and velocity, while a Dirichlet boundary condition is employed for the pressure. The channel walls are set to a constant temperature of 300~K and a zero-gradient boundary condition for the species is assumed (inert wall).
    The flame anchor is located inside the logarithmic region of the turbulent boundary layer at a non-dimensional distance of $z^+ \approx 100$.
    Finally, in the lateral direction, periodic boundary conditions are employed. The reaction mechanism presented by Stagni et al.~\cite{Stagni2021} is used to describe the chemical source terms, and the species diffusion coefficients are modelled using a mixture-averaged transport approach~\cite{Curtiss1949, Coffee1981}. 
\end{itemize}
For the simulations, the finite volume method code OpenFOAM is used with an in-house solver~\cite{Zirwes2018, Zirwes2018improved} based on finite-rate chemistry that was validated as being suitable for DNS-like simulations~\cite{Zirwes2018, Zirwes2019}. For the temporal discretization a second-order implicit backward scheme is employed, while the spatial discretization is of fourth-order.


\section{Results and discussion} \addvspace{10pt}
\subsection{Global flame characteristics} 
\begin{figure}[!htb]
\centering
\includegraphics[width=192pt]{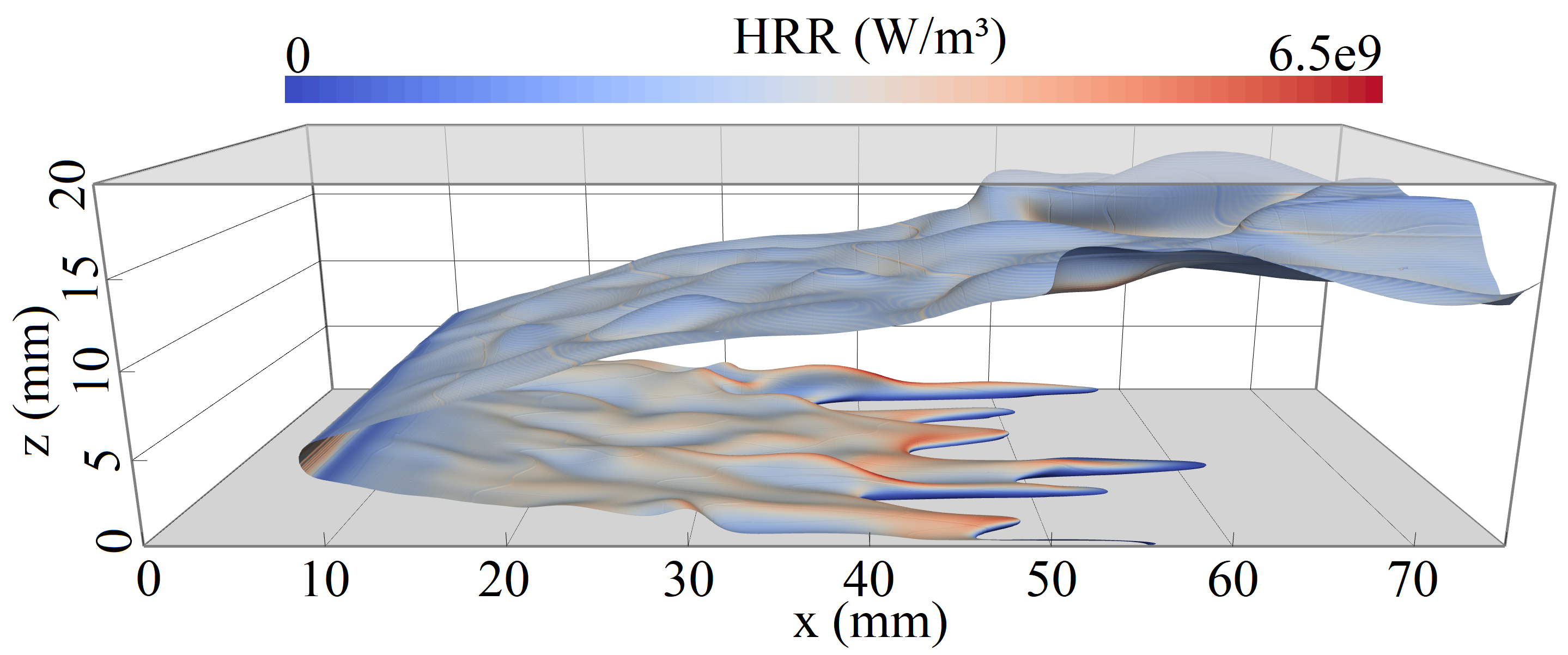}
\vspace{2 pt}
\caption{Instantaneous flame front of the reactive flow simulation represented by the $Y_\mathrm{CO_2}=0.04$ iso-surface and colored by the HRR.}
\label{fig:snapshot}
\end{figure}

\begin{figure}[!htb]
\centering
\includegraphics[width=192pt]{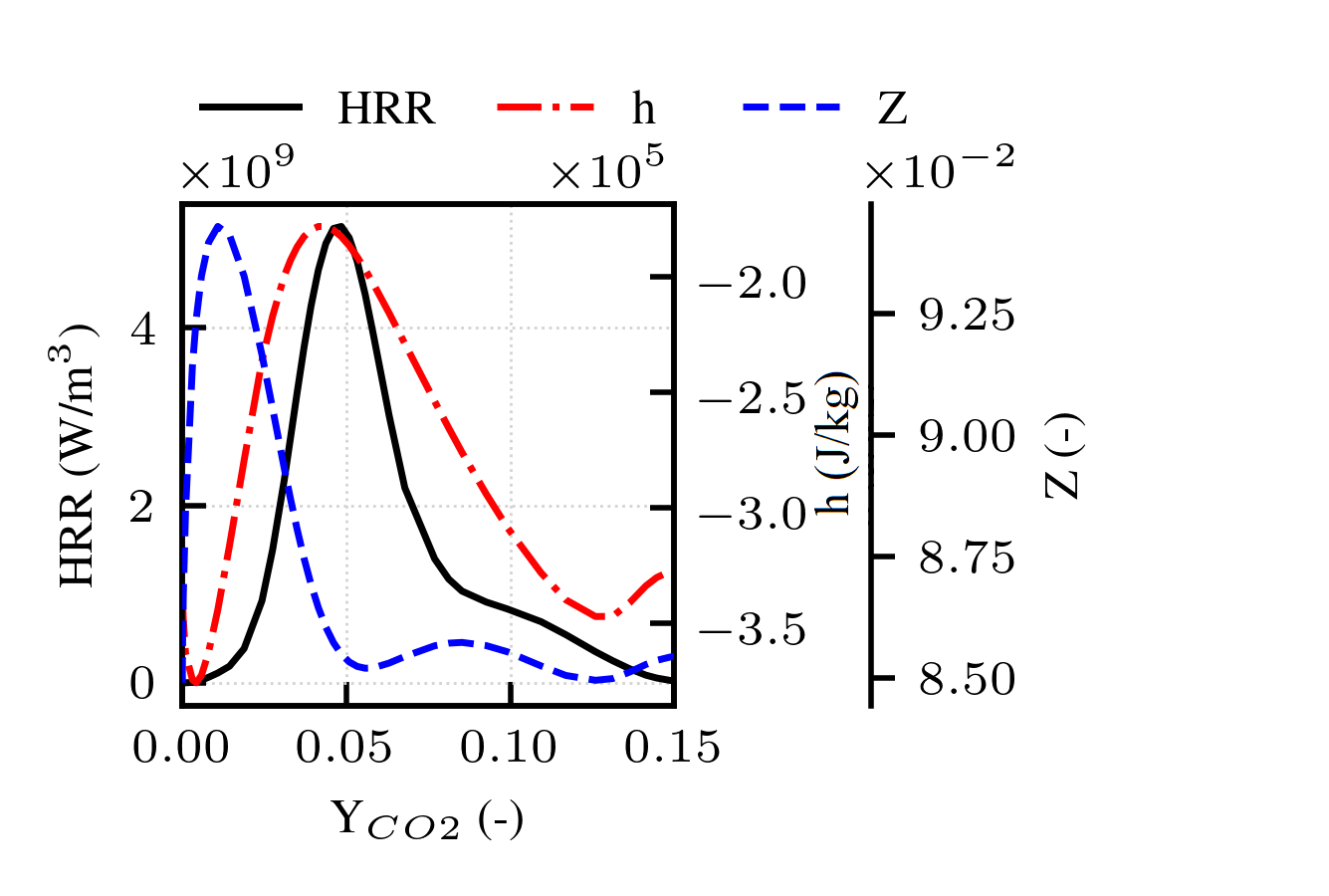}
\caption{Profiles of HRR, h and Z of a one-dimensional, laminar, premixed, unstretched, adiabatic, freely-propagating flame.}
\label{fig:case_1D}
\end{figure}

\addvspace{10pt}
Figure~\ref{fig:snapshot} shows the instantaneous flame front, depicted by the $Y_\mathrm{CO_2}=0.04$ iso-surface and colored to indicate the HRR. This iso-surface is chosen since it is a good approximation of the main reaction front of the flame both in the core flow and the near-wall region.
The iso-surface displays the local variations of HRR and the distinct finger-like features of the flame in the flame wall interaction zone (bottom wall). Note that the lateral direction is statistically independent and can therefore be understood as multiple realizations of the temporal flame evolution.
In the subsequent discussion, all analyses are restricted to the lower flame branch that is affected by enthalpy losses at the walls.
Figure~\ref{fig:case_1D} presents the profiles of the HRR, Bilger's mixture fraction $Z$~\cite{Bilger1990} and absolute enthalpy $h$ in a one-dimensional unstretched laminar premixed freely-propagating flame corresponding to the three-dimensional simulation ($T=300~\mathrm{K}$, $\phi=0.83$) are shown over the $\mathrm{CO_2}$ mass fraction. $\mathrm{CO_2}$ was identified as a suitable species to define the reaction progress in laminar and turbulent DME/air flames~\cite{Zentgraf2022} and it is used for this purpose in this work. In the laminar unstretched flame, differential diffusion leads to a mixture inhomogeneity, which can be seen from the non-uniform profiles of the mixture fraction and absolute enthalpy. 
Furthermore, the mixture fraction and heat release rate profiles are not aligned; instead, the heat release rate peak is shifted towards higher values of $Y_\mathrm{CO_2}$ and thus higher reaction progress.

\begin{figure*}[!htb]
\centering
\includegraphics[scale=1.0]{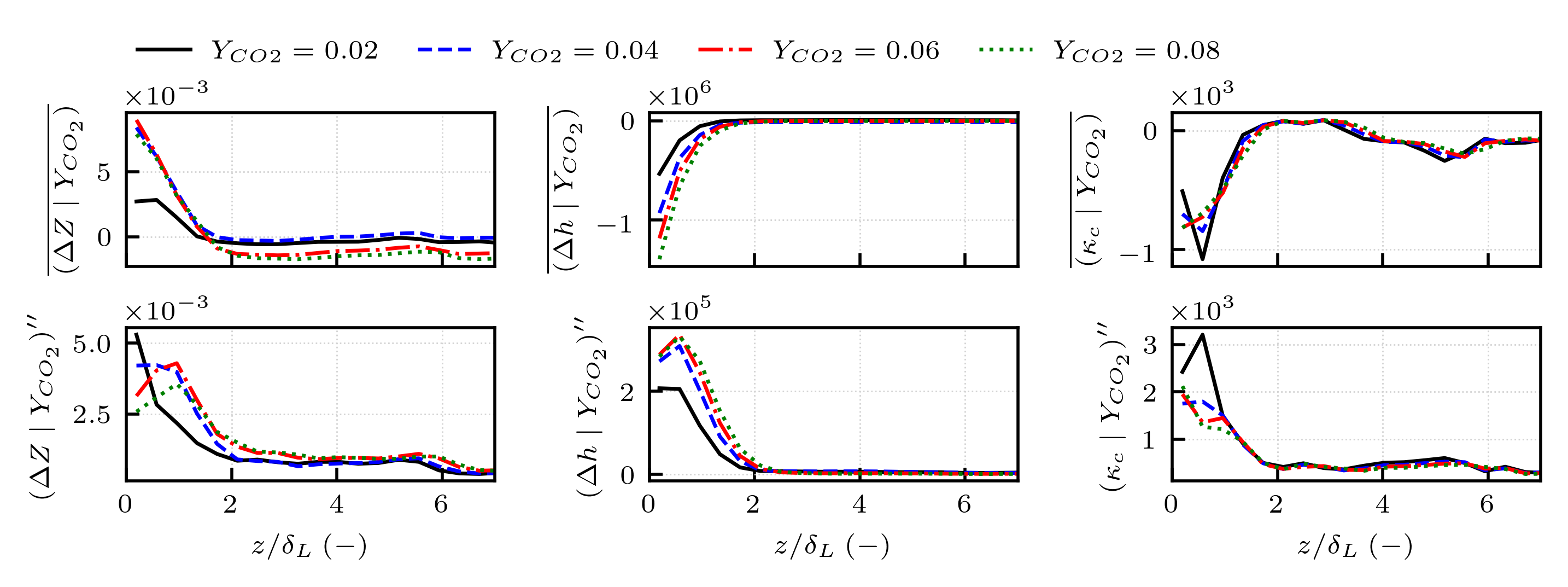}
\vspace{0 pt}
\caption{Values of mean (top) and standard deviation (bottom) of mixture fraction deficit (left), enthalpy deficit (mid) and curvature (right) conditioned on the $\mathrm{CO_2}$ mass fraction over the wall distance.}
\label{fig:cond_means_std}
\end{figure*}

In the three-dimensional reactive case, turbulence induces stretch on the flame, causing the flame front to wrinkle. Additionally, in the vicinity of the wall the flame is affected by heat losses. Both effects, turbulence and heat losses, lead to a deviation of the thermochemical states compared with those of a one-dimensional laminar flame. 
The influence of the wall on the mixture shift is analyzed based on the mixture fraction. To distinguish between effects of turbulence and heat loss and the intrinsic mixture shift due to differential diffusion also present in a one-dimensional flame (see Fig.~\ref{fig:case_1D}), a mixture fraction deficit $\Delta Z$ and an enthalpy deficit $\Delta h$ are defined as
\begin{align}
    \Delta Z \left( Y_\mathrm{CO_2} \right) &= Z_\mathrm{3D} \left( Y_\mathrm{CO_2} \right) - Z_\mathrm{1D} \left( Y_\mathrm{CO_2} \right) \ , \\
    \Delta h \left( Y_\mathrm{CO_2} \right) &= h_\mathrm{3D} \left( Y_\mathrm{CO_2} \right) - h_\mathrm{1D} \left( Y_\mathrm{CO_2} \right)\ , 
\end{align}
with the indices 3D and 1D referring to the three-dimensional reactive simulation and the one-dimensional freely-propagating flame, respectively.
Figure~\ref{fig:cond_means_std} shows the conditional means and standard deviations of the mixture fraction deficit $\Delta Z$, the enthalpy deficit $\Delta h$ and the curvature $\kappa_c$ for a given value of $Y_\mathrm{CO_2}$ over the normalized wall distance $z/\delta_L$.
The flame curvature is defined as $\kappa_c = \nabla \cdot \textbf{n}$ with $\textbf{n}$ being the flame surface normal vector. The influence of the reaction progress is shown for different values of $Y_\mathrm{CO_2}$.

First, in the region $z/\delta_L\geq2$, the core flow unaffected by enthalpy losses is analyzed, before the near-wall region at $z/\delta_L<2$ is discussed. In the core flow, the average flame behaviour is similar to that observed in the one-dimensional laminar flame, indicating conditional means around zero. 
Only small deviations from the zero mean can be observed in the mixture fraction for larger values of $Y_\mathrm{CO_2}$, while $\kappa_c$ shows differences above $z/\delta_L=4$. 
Furthermore, Fig.~\ref{fig:cond_means_std} indicates that while the standard deviation of $\Delta h$ remains zero in the core flow region, $\Delta Z$ and $\kappa_c$ exhibit non-zero standard deviations. They follow a similar trend, indicating a dependency of $\Delta Z$ on $\kappa_c$.

At a wall distance of $z/\delta_L < 2$, the flame is affected by enthalpy losses to the wall, reflected in the decreasing mean and increasing standard deviation of $\Delta h$ at a decreasing wall distance. Due to the higher temperature gradients between the wall and the hot burnt mixture with increasing reaction progress, the enthalpy loss to the wall increases.
Additionally, the curvature $\kappa_c$ decreases and its standard deviation increases towards the wall. $\Delta Z$, on the other hand, shows an increasing mean and standard deviation in the regions closer to the wall. This trend in $\Delta Z$ is less pronounced in the fresh gas region ($Y_\mathrm{CO_2}=0.02$). 
The observed trends of $\kappa_c$ in the near-wall region in the simulation is in contradiction to the experimental investigations by Kosaka et al.~\cite{Kosaka2020}, who observed positive mean curvature values close to the wall. 
These differences could have multiple reasons. Even though the thermochemical states observed in the experimental and numerical configurations are comparable~\cite{Steinhausen2021, Zentgraf2021}, the configurations are not identical, i.e. different flow configuration (Couette flow vs channel flow), a different flame angle to the wall and different Reynolds numbers. These differences might lead to different turbulent statistics in the two flow configurations. Further, the evaluation of curvature in the experiments is limited to the two-dimensional measurement domain, while three-dimensional data are available in the simulation. Hence, the definitions of the evaluated curvature in the experiment and simulation deviate.
FWI is a very local process. A previous study~\cite{Steinhausen2021} showed that events deviating considerably from the mean values occur in turbulent FWI. Therefore, the characteristics seen in the global trends are investigated locally in the core flow and the region of FWI.

\subsection{Local heat release rate and curvature during flame-wall interaction} \addvspace{10pt}

\begin{figure}[!htb]
\centering
\includegraphics[width=192pt]{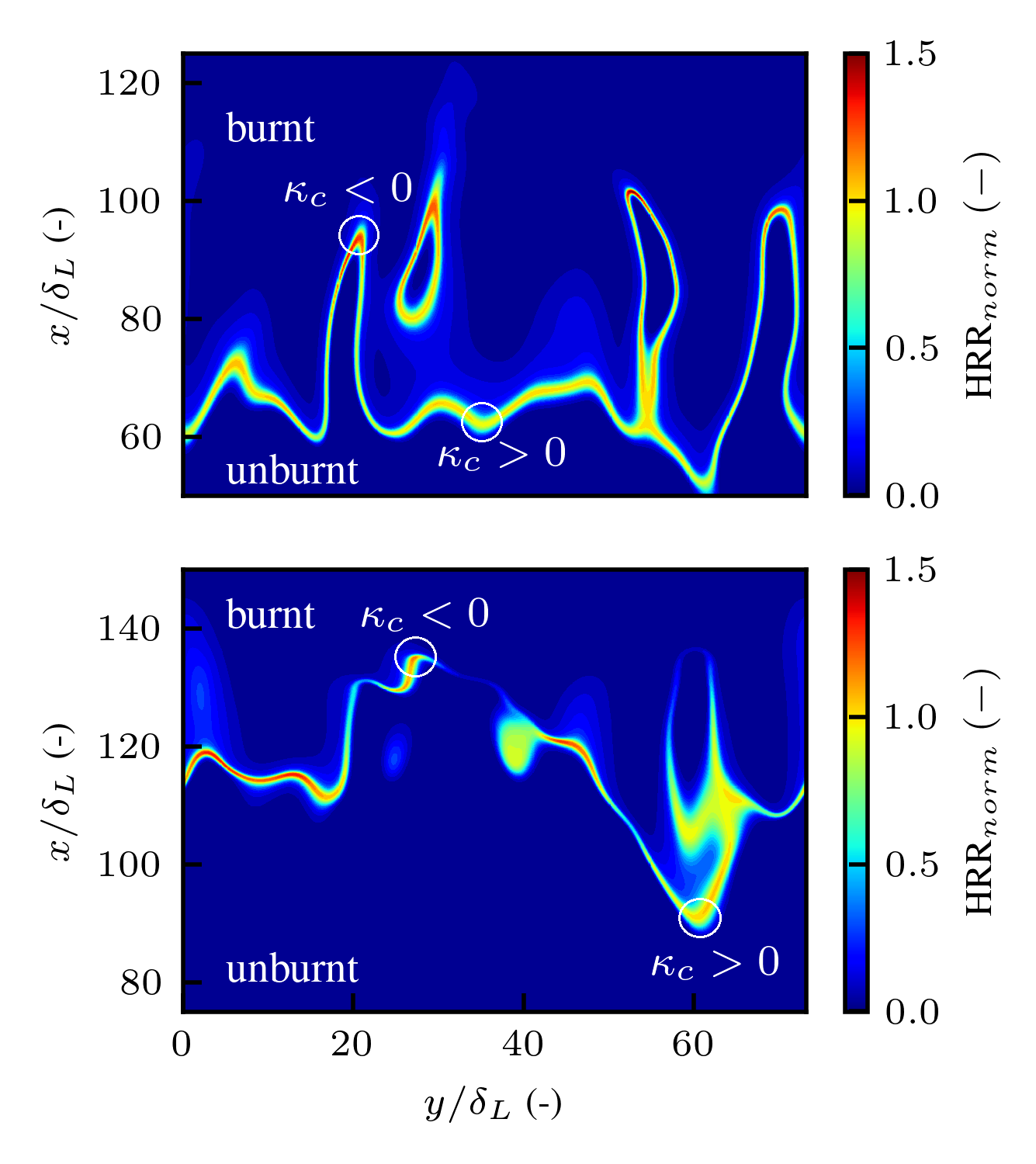}
\caption{Contour plots of the normalized HRR in the wall-parallel plane at distances of $z/\delta_L=4$ (top) and $z/\delta_L=0.5$ (bottom). Regions of negative and positive curvature are indicated by the circles and the captions $\kappa_c < 0$ and $\kappa_c > 0$, respectively.}
\label{fig:slices_hrr}
\end{figure}

\begin{figure*}[!htb]
\centering
\includegraphics[scale=1.0]{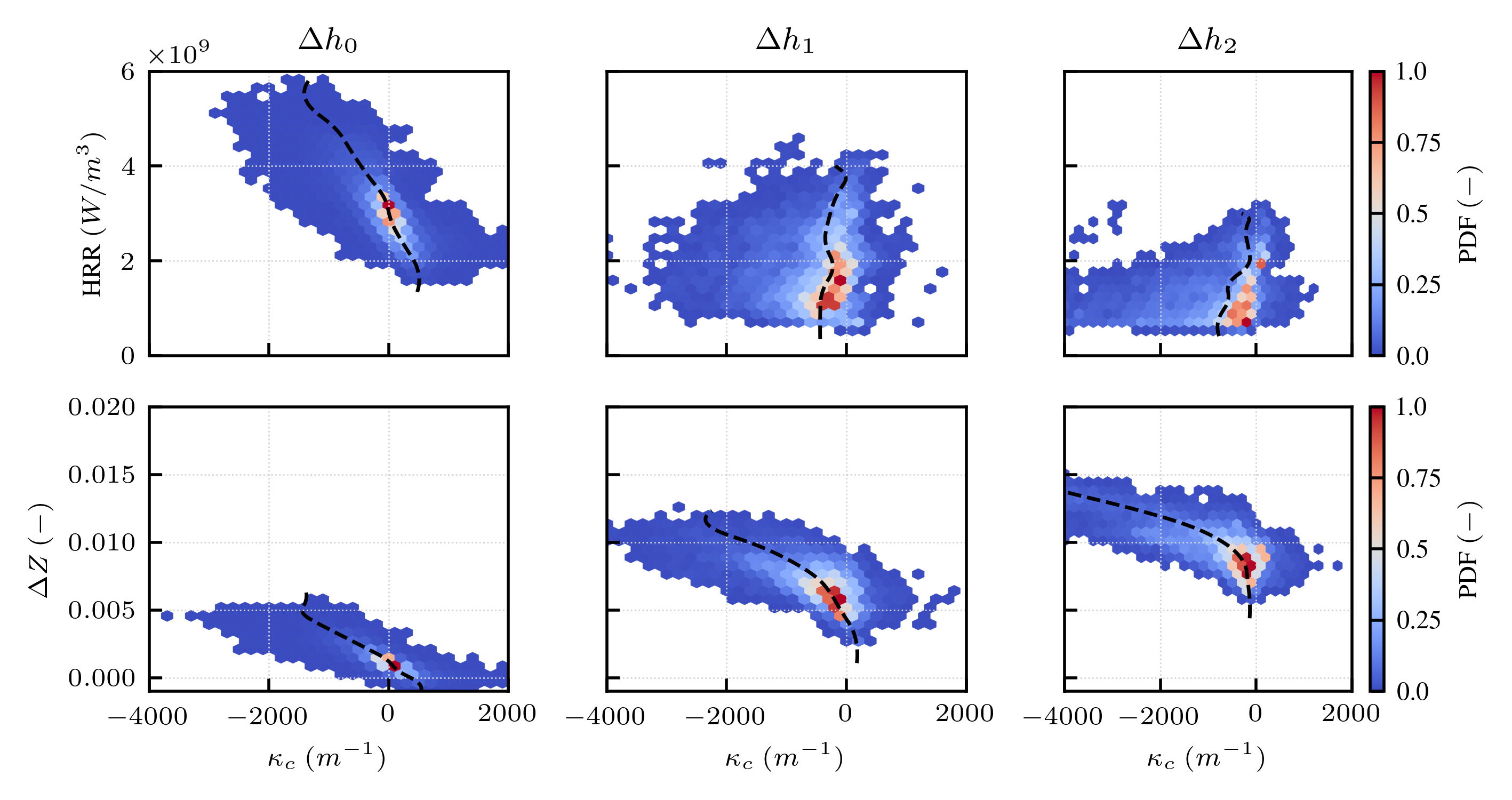}
\vspace{0 pt}
\caption{Joint distributions of $\kappa_c$ and HRR (top) and $\Delta Z$ (bottom) at different levels of enthalpy deficit ($\Delta h_0 = 0~\text{J/kg}$, $\Delta h_1 = -1 \cdot 10^5~\text{J/kg}$, $\Delta h_2 = -3 \cdot 10^5~\text{J/kg}$) at $Y_{CO_2}=0.04$. Mean values are indicated by the dashed line.}
\label{fig:hrr_dz_curv}
\end{figure*}

The effect of curvature on heat release rates is investigated in regions with and without heat loss to the wall.
For a Lewis number above one, HRR and curvature are negatively correlated in weakly curved, adiabatic, steady flames~\cite{lieuwen_2012}.
Without heat loss and the flame only weakly affected by turbulence, it is expected that the flame still shows a similar behaviour as indicated by other numerical studies~\cite{Cecere2016, Bisetti2009}.
Figure~\ref{fig:slices_hrr} shows two contour plots of the HRR normalized by the maximum HRR value of a one-dimensional premixed unstretched adiabatic freely-propagating laminar flame in the wall-parallel plane in the core flow ($z/\delta_L=4$, top) and the near-wall region affected by enthalpy losses to the wall ($z/\delta_L=1$, bottom).
To illustrate the effects on the local HRR in two dimensions, regions of negative $\kappa_c < 0$ and positive curvature $\kappa_c > 0$ are highlighted inside circles. Note that these curvature values are based on the three-dimensional data. In the core flow ($z/\delta_L=4$), the negative curvature is associated with an increase in the local HRR while positive curvature leads to a decrease in the local HRR. This result is in accordance with the experimental findings in Kosaka et al.~\cite{Kosaka2020}.
In the near-wall region affected by heat losses ($z/\delta_L=0.5$, bottom) the regions of negative and positive curvature are no longer clearly associated with variations in the local HRR and both show increased HRR.
This reveals the importance of evaluating the combined effects of heat loss and curvature in FWI.

The combined effects of heat loss and curvature on HRR and $\Delta Z$ are further analyzed in Fig.~\ref{fig:hrr_dz_curv} based on the joint probability density function (PDF) of the local HRR and $\kappa_c$ (top), and $\Delta Z$ and $\kappa_c$ (bottom) at separate levels of $\Delta h$. 
Values are evaluated on the representative $Y_\mathrm{CO_2}=0.04$ iso-surfaces extracted from five time instances and sampled in the intervals of $\Delta h_0 = 0~\text{J/kg}$, $\Delta h_1 = -1 \cdot 10^5\pm 5\%~\text{J/kg}$, $\Delta h_2 = -3 \cdot 10^5\pm 5\%~\text{J/kg}$.
Without heat losses ($\Delta h_0$), the expected behaviour is observed: negative curvatures are associated with an increase in the local HRR while positive curvatures occur with a decreased HRR.

For $\Delta Z$, an identical correlation is observed that is also expected due to the different diffusivities of fuel and oxidizer species.
In the cases of medium and high heat losses ($\Delta h_1$ and $\Delta h_2$), decreased HRR and increased $\Delta Z$ are observed compared to $\Delta h_0$. The correlation between HRR and $\kappa_c$ is inverted, with negative curvatures associated with a decrease in the local HRR. 
For $\Delta Z$ and $\kappa_c$, the previous trend continues and no inversion occurs. 
These trends are more pronounced for high heat losses compared to the medium case.
To summarize, differential diffusion effects on the mixture fraction field increase in the near-wall region, corresponding to higher values of $\Delta Z$ and an unaffected sensitivity to $\kappa_c$. In contrast to this, the HRR is strongly reduced by heat losses and exhibits an inverted sensitivity to $\kappa_c$. During laminar SWQ of methane/air flames, similar behaviour is reported by Zhang et al.~\cite{Zhang2021}, who observe a change in the sign of the correlations between local flame speeds and normalized flame curvature in the region of FWI.

\subsection{Dependency of local heat release rate and curvature on quenching events} 
\begin{figure}[!htb]
\centering
\includegraphics[width=192pt]{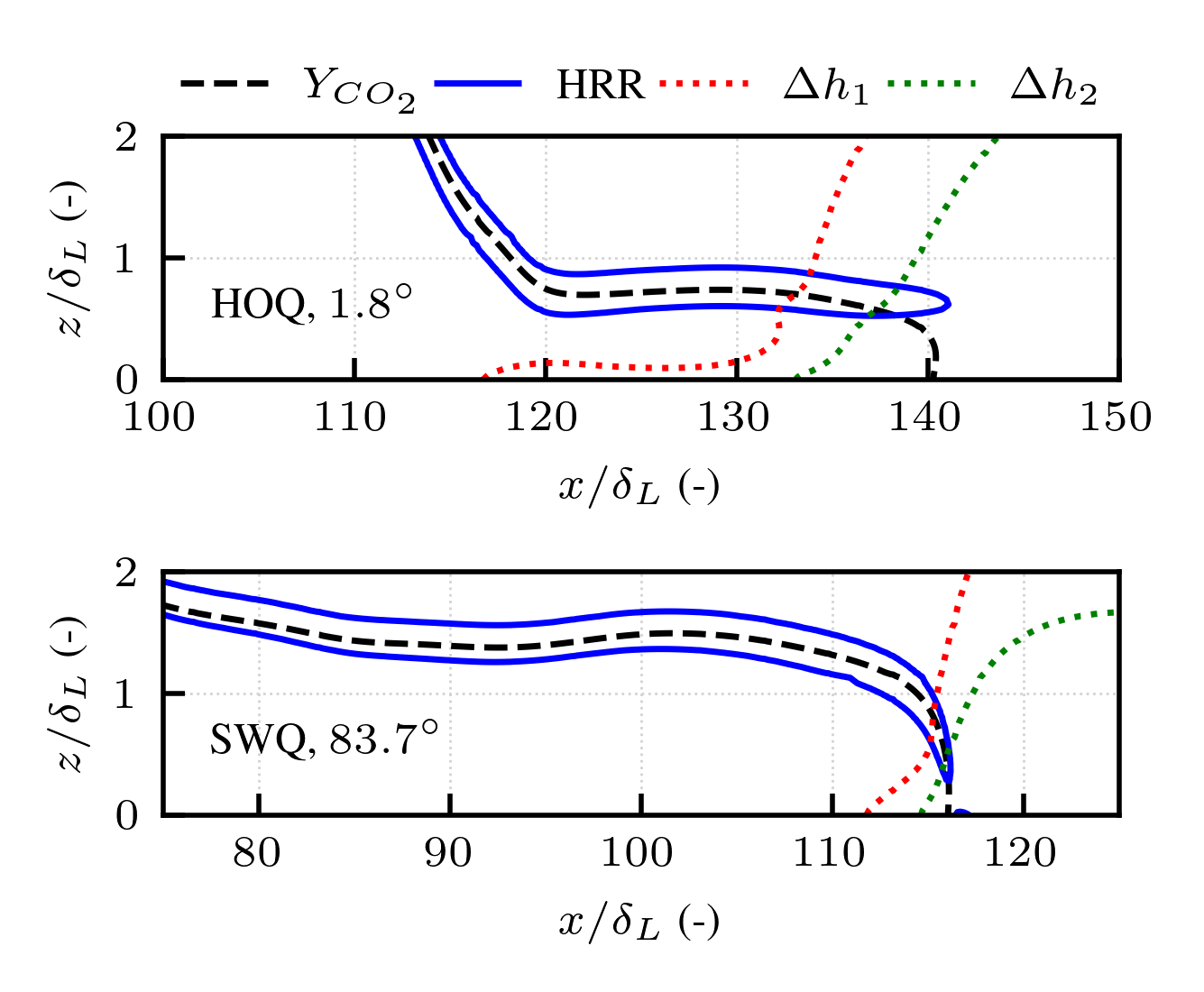}
\caption{Instances of the two quenching scenarios in a lateral slice: HOQ (top), SWQ (bottom). Iso-contours of $Y_\mathrm{CO_2}=0.04$ (black dashed line) and $\mathrm{HRR}=0.1 \cdot \mathrm{HRR}_\text{max}$ (blue solid line) indicate the flame surface. Iso-contours of enthalpy deficit are shown at $\Delta h_1= -1 \cdot 10^5~\text{J/kg}$ (red dotted line) and $\Delta h_2= -3 \cdot 10^5~\text{J/kg}$ (green dotted line).}
\label{fig:quenching_scenarios}
\end{figure}

\begin{figure}[!htb]
\centering
\includegraphics[scale=1.0]{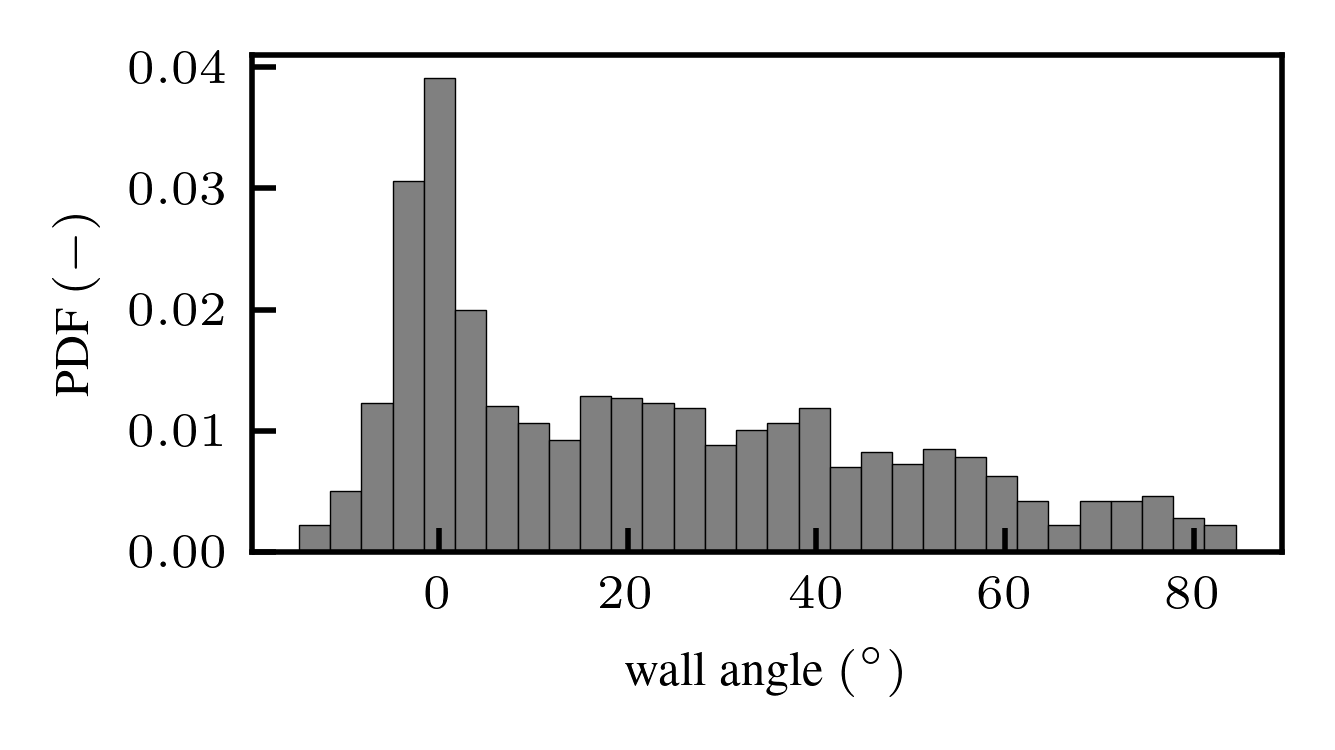}
\caption{Probability distribution of the flame–wall angle at the quenching point.}
\label{fig:flame_angle}
\end{figure}

\addvspace{10pt}
To better understand the sensitivity of the HRR-curvature correlations in the FWI zone, the influence of flame quenching on curvatures and HRR is investigated. Turbulent SWQ quenching occurs in a multitude of flame orientations, spanning a spectrum from HOQ-like to SWQ-like behaviour~\cite{Heinrich2018, Zentgraf2021}. 
The quenching modes can be classified based on the angle between the flame surface-normal vector and the wall-normal direction at the point of quenching. The flame orientation is validated in the lateral direction such that for each slice in y direction the quenching mode/angle is defined separately. This is in line with other experimental~\cite{Zentgraf2021} and numerical~\cite{Heinrich2018} works in the literature. Figure~\ref{fig:quenching_scenarios} shows two lateral slices, first of a HOQ-like (top) then of a SWQ-like (bottom) quenching scenario. The flame is represented by the contours of $\mathrm{HRR}=0.1 \cdot \mathrm{HRR}_\text{max}$ and $Y_\mathrm{CO_2}=0.04$ as blue solid line and black dashed line, respectively. $\mathrm{HRR}_\text{max}$ denotes the global maximum of the observed HRR values.
The intersection of the two contours mark the quenching point of the flame. Further, two iso-contours of $\Delta h$ are indicated as red and green dotted lines. Looking at the medium and high enthalpy deficits ($\Delta h_1$ and $\Delta h_2$), one distinct difference between HOQ- and SWQ-like quenching becomes apparent. In the HOQ-like scenario, the flame burns parallel to the wall (at a relatively low wall distance). Hence, the region of enthalpy loss is distributed over a wide region at the wall in stream-wise direction $x$. In the SWQ-like scenario, however, the flame exhibits a more distinct quenching point, and the region of heat loss to the wall is limited to a small area in the stream-wise direction where the flame tip tilts towards the wall. This quenching shape is similar to that observed in a laminar SWQ flame.
Figure~\ref{fig:flame_angle} shows the distribution of the flame angle evaluated over five time steps and further averaged over the statistically independent lateral direction. In the distribution, angles between $0^{\circ}$ and $90^{\circ}$ are present and a distinct peak around $0^{\circ}$ can be observed that flattens towards $90^{\circ}$. In the analysis below, the flame angles are classified as HOQ- and SWQ-like events. Following Heinrich et al.~\cite{Heinrich2018}, all flame angles smaller than $2^{\circ}$ are classified as HOQ-like events exhibiting a reaction front parallel to the wall, while the other cases are assumed to exhibit SWQ-like behaviour with a distinct quenching point (see Fig.~\ref{fig:quenching_scenarios}). 

\begin{figure}[!htb]
\centering
\includegraphics[scale=1.0]{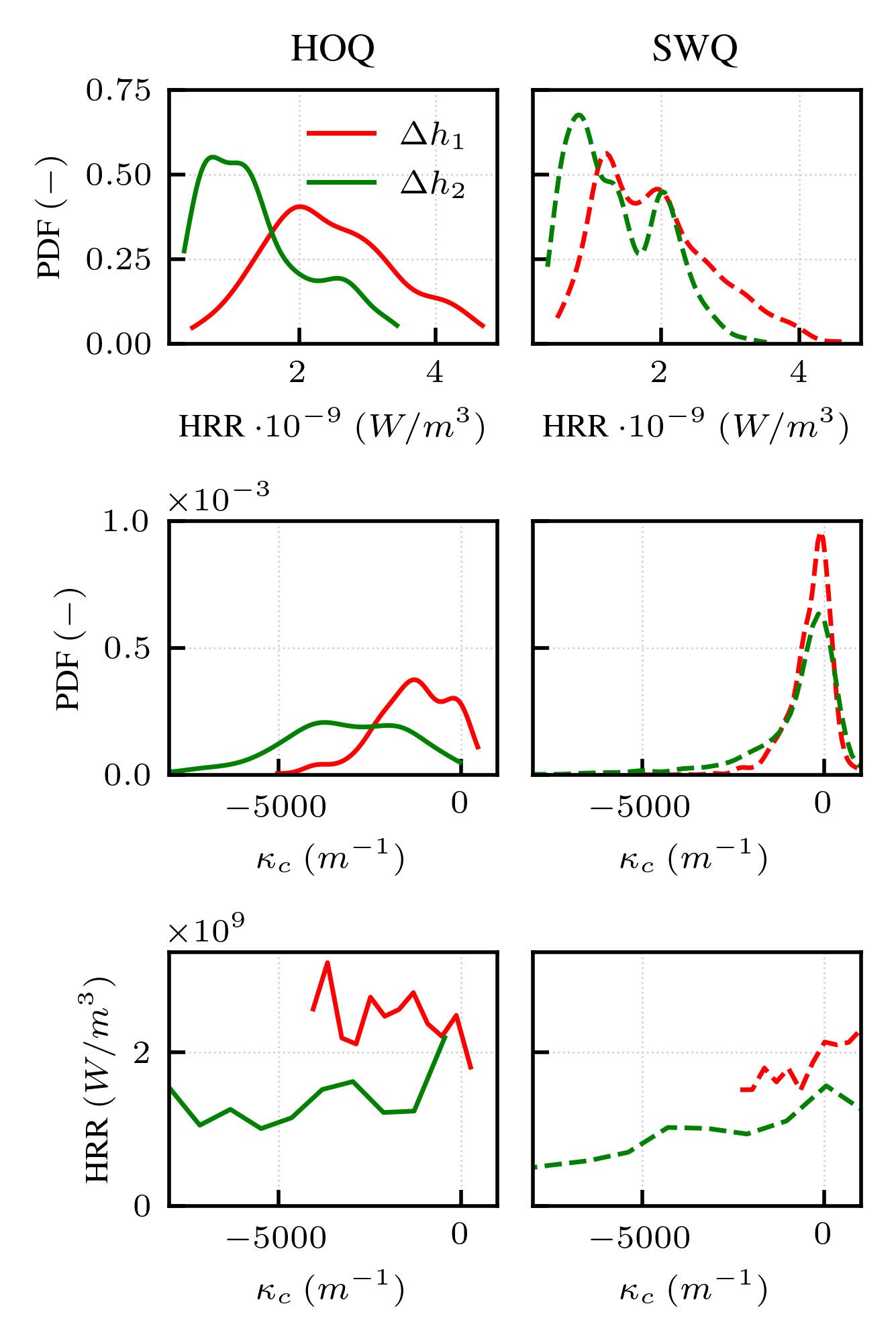}
\caption{PDFs of local HRR (top) and $\kappa_c$ (middle), mean values of HRR conditioned on $\Delta h$ over $\kappa_c$ (bottom) in HOQ (left) and SWQ (right) events.}
\label{fig:hoq_swq_hrr}
\end{figure}

In this section, differences in the local HRR and curvatures are investigated for HOQ and SWQ events.
Figure~\ref{fig:hoq_swq_hrr} shows the PDFs of the local HRR (top) and curvature $\kappa_c$ (middle), and the mean values of the HRR over $\kappa_c$ (bottom). 
The distributions are evaluated  separately for each quenching scenario – HOQ (left, solid lines) and SWQ  (right, dashed lines) – at the two levels of enthalpy deficit $\Delta h_1$ and $\Delta h_2$. 
Note that the data is extracted on the $Y_\mathrm{CO_2}=0.04$ iso-surface to allow the flame curvature to be assessed.
In HOQ, the HRR decreases with increasing heat loss. This is reflected in the shift of the PDF towards lower HRR values for a higher enthalpy deficit. In the same manner, the distribution of $\kappa_c$ moves to more negative values with increasing heat loss. The correlation between HRR and $\kappa_c$ shows an inversion of its signs from negative to positive with increasing heat loss.
In SWQ, a similar shift towards lower HRR with increasing heat loss is observed. 
The distributions of $\kappa_c$ display a distinct peak around zero and for increasing heat loss the distribution is broadened towards negative values. The correlations between HRR and $\kappa_c$ are positive for both enthalpy deficits.
Compared to HOQ, SWQ is associated with lower HRR. The distributions of curvature differ significantly. HOQ is associated with higher negative values of curvature and qualitatively different shapes of the distributions compared to SWQ. Correlations between HRR and $\kappa_c$ are similar for HOQ and SWQ at $\Delta h_2$, while they are negatively correlated in HOQ and positively correlated in SWQ at $\Delta h_1$.
These differences may explain the slightly higher HRR observed in HOQ compared to SWQ at $\Delta h_1$.
Further, this might indicate that the onset of the transition from a negative correlation between HRR and $\kappa_c$ to a positive correlation during FWI is sensitive to the local quenching mode.

\section{Summary and conclusions} \addvspace{10pt}

In this work, a fully resolved numerical simulation of a turbulent premixed DME/air flame in a generic SWQ configuration is conducted taking into consideration finite-rate chemistry and mixture-averaged diffusion. The combined effects of curvature and heat loss on the local HRR and mixture fraction are assessed using the detailed insights from the numerical data. First, the mean flame behaviour is assessed both in the core flow and the near-wall region, which is considered to be limited to a wall distance smaller than twice the laminar flame thickness. In this region, the conditional mean and averages of the mixture fraction and curvature are strongly affected by the wall. Further, the correlation of the local HRR and curvature is different in that region. 
In the core flow that is unaffected by enthalpy losses to the wall, the local HRR and curvature show a negative correlation, i.e. an increase in the local HRR with negative curvature, which is in accordance with the literature. In the near-wall region, however, the correlation is inverted. The mixture fraction shows a negative correlation in both regions and only the mean value increases with decreasing enthalpy loss to the wall. 
Finally, the impact of the quenching event on the flame behaviour is assessed. To this end, the quenching is classified as head-on quenching-like (with reaction front parallel to the wall) and side-wall quenching-like behaviour, based on the flame angle to the wall. In the head-on quenching-like scenario, the flame is quenched at a lower wall distance and is associated with higher negative values of curvature compared to the side-wall quenching-like event. Further, the correlation of the local heat release and curvature is sensitive to the quenching mode for medium heat losses to the wall. 
In contrast, at high heat losses or at enthalpy levels close to the fully quenched flame, the correlation between HRR and curvature is not dependent on the quenching mode.
 
The findings discussed above indicate that there is a strong correlation between the turbulent flame dynamics and flame–wall interaction that significantly affects the flame behaviour close to the (cold) walls.
The correlation of HRR on curvature was found to be sensitive to heat losses, illustrating the importance of suitable turbulent combustion models when simulating near-wall flows. While the current work investigated the effects of the curvature, future works should evaluate the combined effects of strain and curvature in turbulent DME/air flames during FWI.

\section*{Acknowledgments} \addvspace{10pt}
The research leading to these results has received funding from the European Union’s Horizon 2020 research and innovation programme under the Center of Excellence in Combustion project, grant agreement No 952181, from the Deutsche Forschungsgemeinschaft (DFG, German Research Foundation) -- Project Number 237267381 -- TRR 150, and from the Federal Ministry of Education and Research (BMBF) and the state of Hesse as part of the NHR Program. 
Calculations for this research were conducted on the Lichtenberg high-performance computer of the TU Darmstadt.

\newpage


\bibliography{publication.bbl}
\bibliographystyle{unsrtnat_mod}

\end{document}